%% file: HIG-14-013_temp.tex
\pdfoutput=1

\documentclass[11pt,twoside,a4paper,cmspaper,final,collab]{cms-tdr}
\def\svnVersion{303688}\def\cmsCernNo{2013-037}\def\cmsCernDate{\today}\def\cmsMessage{Published in Physics Letters B as \href{http://dx.doi.org/10.1016/j.physletb.2015.08.047}{\doi{10.1016/j.physletb.2015.08.047.}}}

\begin{document}\cmsNoteHeader{HIG-14-013}

\hyphenation{had-ron-i-za-tion}
\hyphenation{cal-or-i-me-ter}
\hyphenation{de-vices}
\RCS$Revision: 300501 $
\RCS$HeadURL: svn+ssh://svn.cern.ch/reps/tdr2/papers/HIG-14-013/trunk/HIG-14-013.tex $
\RCS$Id: HIG-14-013.tex 300501 2015-08-18 14:30:46Z alverson $

\newlength\cmsFigWidth
\ifthenelse{\boolean{cms@external}}{\setlength\cmsFigWidth{0.98\columnwidth}}{\setlength\cmsFigWidth{0.6\textwidth}}
\newlength\cmsFigWidthThree
\ifthenelse{\boolean{cms@external}}{\setlength\cmsFigWidthThree{0.48\textwidth}}{\setlength\cmsFigWidthThree{0.48\textwidth}}
\ifthenelse{\boolean{cms@external}}{\providecommand{\cmsLeft}{top\xspace}}{\providecommand{\cmsLeft}{left\xspace}}
\ifthenelse{\boolean{cms@external}}{\providecommand{\cmsRight}{bottom\xspace}}{\providecommand{\cmsRight}{right\xspace}}
\newcommand{\XHH}{\ensuremath{X \to \PH\PH}\xspace}
\newcommand{\XHbbHbb}{\ensuremath{X \to \PH\PH \to \bbbar\bbbar}\xspace}
\newcommand{\ppXHbbHbb}{\ensuremath{\Pp\Pp \to X \to\PH\PH \to \bbbar\bbbar}\xspace}
\providecommand{\cPX}{\ensuremath{X}\xspace}
\newcolumntype{.}{D{.}{.}{-1}}

\cmsNoteHeader{HIG-14-013}
\title{Search for resonant pair production of Higgs bosons decaying to two bottom quark-antiquark pairs in proton-proton collisions at 8\TeV}

\date{\today}

\abstract{
A model-independent search for a narrow resonance produced in proton-proton collisions at $\sqrt{s} = 8$\TeV and decaying to a pair of 125\GeV Higgs bosons that in turn each decays into a bottom quark-antiquark pair is performed by the CMS experiment at the LHC. The analyzed data correspond to an integrated luminosity of 17.9\fbinv. No evidence for a signal is observed. Upper limits at a 95\% confidence level on the production cross section for such a resonance, in the mass range from 270 to 1100\GeV, are reported. Using these results, a radion with decay constant of 1\TeV and mass from 300 to 1100\GeV, and a Kaluza--Klein graviton with mass from 380 to 830\GeV are excluded at a 95\% confidence level.
}

\hypersetup{%
pdfauthor={CMS Collaboration},%
pdftitle={Search for resonant pair production of Higgs bosons decaying to two bottom quark-antiquark pairs in proton-proton collisions at 8 TeV},%
pdfsubject={CMS},%
pdfkeywords={CMS, physics, Higgs, bottom, di-Higgs, radion, KK graviton, RS1}}

\maketitle

\section{Introduction}

Following the discovery of a Higgs boson (\PH) at the CERN LHC~\cite{Aad20121, Chatrchyan201230, Chatrchyan:2013lba}, with mass around 125\GeV and properties so far consistent with the standard model (SM) of particle physics, it has become important to search for new resonances that decay into pairs of such Higgs bosons. While non-resonant pair production of the Higgs boson is allowed in the SM, the theoretical production cross section is approximately 10\unit{fb}~\cite{deFlorian:2013jea} and well beyond the sensitivity of currently acquired data. However, several well-motivated hypotheses of physics beyond the standard model posit narrow-width resonances that decay into pairs of Higgs bosons, and could be produced with large enough cross sections to be probed with existing data. The radion~\cite{GoldbergerWise} and Kaluza--Klein (KK) gravitons in the  Randall--Sundrum (RS1)~\cite{RandallSundrum} model of warped extra dimensions are examples of such resonances~\cite{radion}.

This letter reports the results of a model-independent search for the resonant pair production of Higgs bosons. The search for the narrow width resonance, denoted by $\cPX$, is performed in the 270--1100\GeV mass range. Data from proton-proton collisions at the LHC and recorded by the CMS experiment corresponding to an integrated luminosity of $17.9 \pm 0.5$\fbinv at $\sqrt{s} = 8$\TeV is used. We perform this search for the case where both Higgs bosons decay into bottom quark-antiquark pairs (\bbbar)~\cite{Chatrchyan:2013zna}. The main challenge of this search is to distinguish the signal of four bottom quarks in the final state that hadronize into jets (b jets) from the copious multijet background described by quantum chromodynamics (QCD) in $\Pp\Pp$ collisions. We address this challenge by suitable event selection criteria that include dedicated b-jet identification techniques and a model of the multijet background that is validated in data control regions. Our results may be compared with a search performed by the ATLAS experiment~\cite{Aad:2014yja} that also probes the physics of resonant Higgs boson pair production, albeit in the channel where one Higgs boson decays to bottom quarks and the other decays to photons.

\section{Detector and Event Reconstruction}

A detailed description of the CMS detector, together with a description of the coordinate system used and the relevant kinematic variables, can be found in Ref.~\cite{Chatrchyan:2008aa}. The central feature of the CMS apparatus is a superconducting solenoid of 6\unit{m} internal diameter that generates an axial magnetic field of 3.8\unit{T}. Within the solenoid volume are a silicon pixel and strip tracker, a lead tungstate crystal electromagnetic calorimeter, and a brass and scintillator hadron calorimeter. Muons are detected and their properties measured in gas-ionization detectors embedded in a steel flux-return yoke outside the solenoid. Jets are reconstructed using the anti-\kt clustering algorithm~\cite{antikt, Cacciari:fastjet1} with a distance parameter of 0.5 applied on the collection of particle candidates reconstructed by the particle-flow (PF) algorithm~\cite{CMS-PAS-PFT-09-001, CMS-PAS-PFT-10-001}. The PF algorithm reconstructs and identifies each individual particle with a combination of information from the various elements of the CMS detector. To mitigate the effect of additional particles that do not originate from the hard interaction in jet reconstruction, we subtract charged hadrons that do not arise from the primary vertex associated with the jet from the collection of clustered particles. Further, an average neutral energy density from particles not arising from the primary vertex is evaluated and subtracted from the jets~\cite{Cacciari:2007fd}. Energy corrections for the jets are determined as functions of the jet transverse momentum \pt and pseudorapidity $\abs{\eta}$. Jet identification criteria~\cite{CMS-PAS-JME-13-005} to reject detector noise misidentified as jets, and jets not originating from the hard interaction are also applied.

In order to identify (tag) b jets, we rely on the fact that bottom quarks hadronize into b hadrons which have decay lengths of the order of $c\tau=450\mum$. Thus, their decay products originate from secondary vertices made of tracks that have impact parameters with respect to the primary vertex of a similar scale. The pixel tracker provides an impact parameter resolution of about 15\mum for charged tracks with $\abs{\eta} < 2.4$. To maximize the b-tagging performance of the detector, we combine the output discriminants of several b-tagging algorithms described in Ref.~\cite{Chatrchyan:2012jua} with a trained artificial neural network. This we call the combined multivariate (CMVA) algorithm. In particular, we combine the outputs of the combined secondary vertex (CSV) tagger that uses secondary vertices identified by the inclusive vertex finder (IVF) algorithm~\cite{BPH-10-010}, the jet probability (JP) tagger, and the two soft lepton taggers.

The first level of the trigger, consisting of customized processors, collects data for this analysis using information from the calorimeters and requires two jets to exceed \pt thresholds of 56 or 64\GeV, depending on luminosity conditions. The second level of the trigger, consisting of software algorithms executed on a farm of commercial processors, uses information from the entire detector to reconstruct PF jets, and requires four PF jets with $\abs{\eta} < 2.4$ and $\pt >30$\GeV, of which two jets must have $\pt>80$\GeV. Further, to record signal events and reject background QCD multijet events, two jets are required to be tagged by the CSV b-tagging algorithm implemented at the trigger.

\section{Simulated Samples}

To model the production of a generic narrow-width spin-0 resonance, we use a Monte Carlo simulation of the RS1 radion produced through gluon fusion. The angular distributions of a spin-2 resonance are distinct from those of a spin-0 resonance, and result in different kinematic distributions. Therefore, we evaluate the signal efficiencies for a narrow-width spin-2 resonance from a separate simulation of the first excitation of the KK graviton produced through gluon fusion in the same extra dimension scenario as the radion. The resonance is forced to decay to a pair of Higgs bosons where both Higgs bosons decay to \bbbar. Samples of these signal events, as well as background events from diboson, \PW+jets, \cPZ+jets and top-quark pair production (\ttbar) processes, are generated using the \MADGRAPH 5.1~\cite{Alwall:2011uj} program interfaced with \PYTHIA 6.4~\cite{PYTHIA} for parton showering and hadronization. QCD multijet event samples are simulated with the \PYTHIA 6.4 program. A sample of events where the Higgs boson is produced in association with a \cPZ ~boson is simulated using the \POWHEG event generator~\cite{POWHEG2, POWHEG1, POWHEG3} interfaced with the \HERWIGpp~\cite{Gieseke:2006ga} program for showering and hadronization. We set the \PYTHIA 6.4 parameters for the underlying event to the Z2* tune~\cite{Chatrchyan:2013gfi}. The response of the CMS detector is modeled using \GEANTfour~\cite{GEANT4}.

On average, 21 $\Pp\Pp$ interactions occurred per bunch crossing in the data used in this analysis. Additional simulated $\Pp\Pp$ interactions overlapping with the event of interest were added to the simulated samples to reproduce the distribution of the number of primary vertices per event reconstructed in data.

\section{Event Selection}
\label{sec:EventSelection}

\begin{figure}[tb]
\centering
\includegraphics[width=\cmsFigWidth]{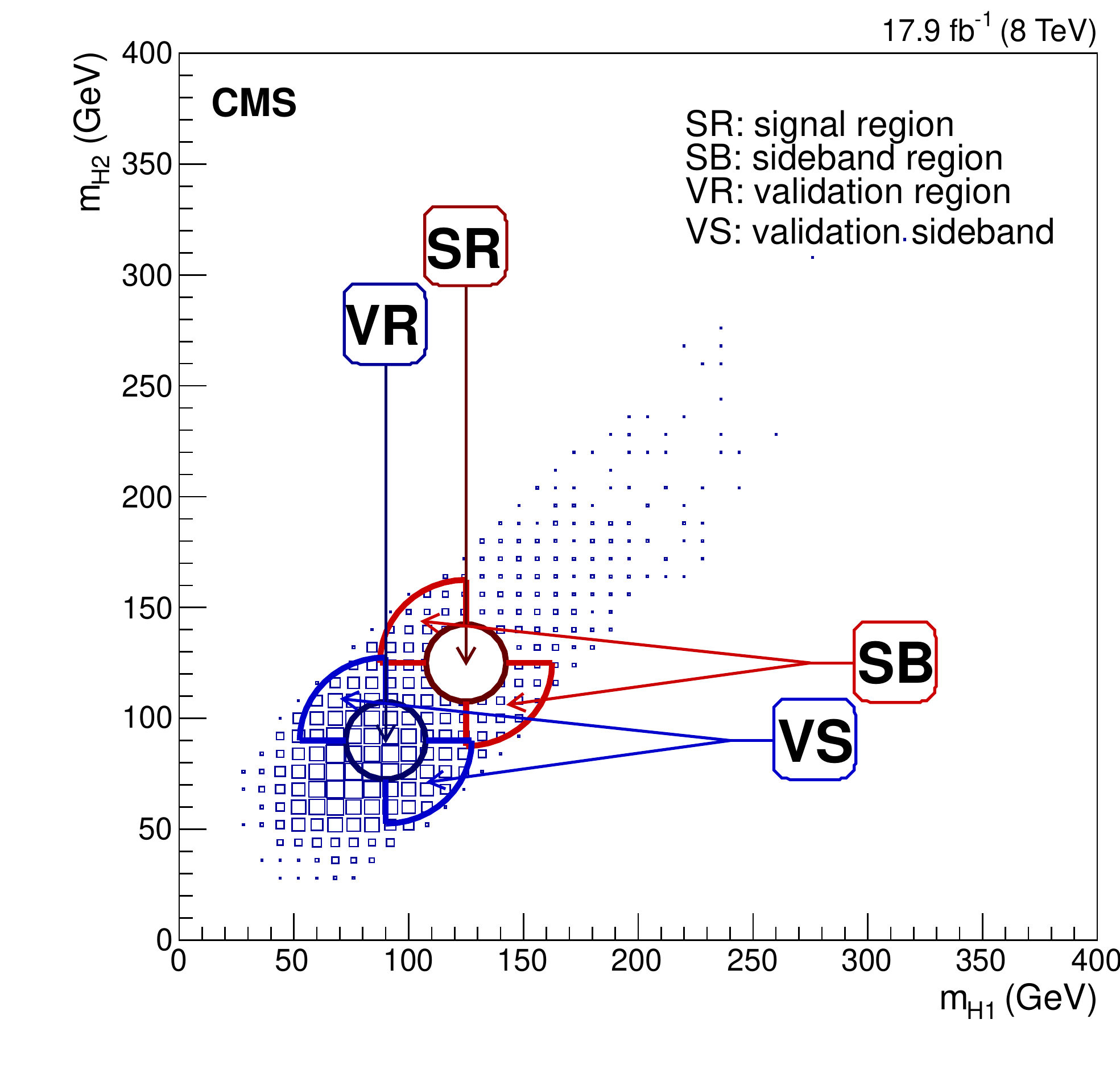}

\caption{Illustration of the SR, SB, VR and VS kinematic regions in the ($m_{{\PH}_1}$, $m_{{\PH}_2}$) plane used to motivate and validate the parametric model for the QCD multijet background. The quantities $m_{{\PH}_1}$ and $m_{{\PH}_2}$ are the two reconstructed Higgs boson masses. The distribution in data events after b-tagging and kinematic selections is shown with the SR blinded.}
\label{fig:Strategy}
\end{figure}

The trigger-level jet \pt thresholds confine our search for a narrow-width \XHbbHbb resonance to masses above 270\GeV. Beyond $m_\cPX \approx$ 800\GeV, the selection efficiency is increasingly limited by the merging of jets from the same Higgs boson, and we curtail this search at 1100\GeV. The kinematic distributions of the decay products vary substantially over this mass range. Therefore, to optimize the search sensitivity, we use different event selection criteria in three main kinematic regions: the low-mass region (LMR) for mass hypotheses from 270 to 450\GeV, the medium-mass region (MMR) for masses from 450 to 730\GeV, and the high-mass region (HMR) for masses from 730 to 1100\GeV.

Event selection begins with the identification of events containing at least four jets in the central region of the detector ($\abs{\eta} < 2.4$) that are b-tagged and have $\pt>40$\GeV. To b-tag a jet, we require it to pass a working point for the CMVA algorithm that maximizes the sensitivity of this search. For jets with $\pt>40$\GeV and $\abs{\eta} < 2.4$ this working point yields a 75\% efficiency for tagging jets originating from b hadrons and a mistagging rate of 3\% for light-flavor jets. For the LMR, we combine these b jets into pairs to create HH candidates such that $\abs{m_\PH - 125\GeV} < 35\GeV$ for each candidate Higgs boson. The mass resolution on the Higgs boson in the LMR is found to be approximately 9\GeV. Selected HH candidates are required to have at least two jets with $\pt>90$\GeV. In the MMR, signal events have large Lorentz factors for the Higgs boson candidates. Therefore, HH candidates for this region are constructed from four jets such that the $\Delta R = \sqrt{\smash[b]{\Delta\eta^2 + \Delta\phi^2}}$ between the jets associated with an \PH candidate remain within 1.5, where $\Delta\eta$ and $\Delta\phi$ are the differences in the pseudorapidities and azimuthal angles of the two jets. For the HMR, we use the same criteria used in the MMR with an additional requirement of $\pt>300$\GeV on one of the \PH candidates to better discriminate signal events from background. In all three regions, in case of multiple HH candidates in an event, the combination with the smallest $|m_{{\PH}_1} - m_{{\PH}_2}|$ is chosen. Having identified the two Higgs boson candidates in each event, we plot their masses, $m_{{\PH}_1}$ and $m_{{\PH}_2}$, on a two-dimensional histogram as shown in Fig.~\ref{fig:Strategy}. ${\PH}_1$ and ${\PH}_2$ are chosen at random from the two reconstructed \PH candidates. As the final selection criterion applied in each of the three mass hypothesis regions, we require events to fall within the signal region (SR) defined as $\sqrt{\smash[b]{\Delta m_{{\PH}_1}^{2} + \Delta m_{{\PH}_2}^{2}}}<17.5$\GeV, where $\Delta m_{{\PH}_{1,2}} = m_{{\PH}_{1,2}} - 125$\GeV.

The efficiencies of these selection criteria for spin-0 and spin-2 resonances at representative masses are shown in Table~\ref{tab:SignalEfficiencies}. The major loss in efficiency for all mass hypotheses comes from the b-tagging requirement for 4 jets. For the 300\GeV mass hypothesis, this is compounded by the trigger inefficiency. The distribution of the aforementioned $\Delta R$ between jets from a single Higgs boson is narrower for the spin-2 resonance, and thus requiring $\Delta R <1.5$ results in a higher efficiency for it.

\section{Signal Modeling}

\begin{table}[tbp]
\topcaption{Efficiencies of the event selection criteria for generic spin-0 and spin-2 resonances decaying to a pair of Higgs bosons in the four b jet final state at representative masses.}
\label{tab:SignalEfficiencies}
\centering
\begin{tabular}{cccccc}
\hline
\multirow{2}{*}{selection eff. (\%)}&\multicolumn{5}{c}{Mass (\GeVns)}\\\cline{2-6}
       & 300 & 500 & 700 & 900 & 1100 \\
\hline
spin-0   & 0.05 & 2.3  & 4.9  & 4.6  & 2.2 \\
spin-2& 0.09 & 3.4  & 6.6  & 5.4  & 2.3 \\
\hline
\end{tabular}

\end{table}

\begin{figure}[tb]
\centering
\includegraphics[width=\cmsFigWidth]{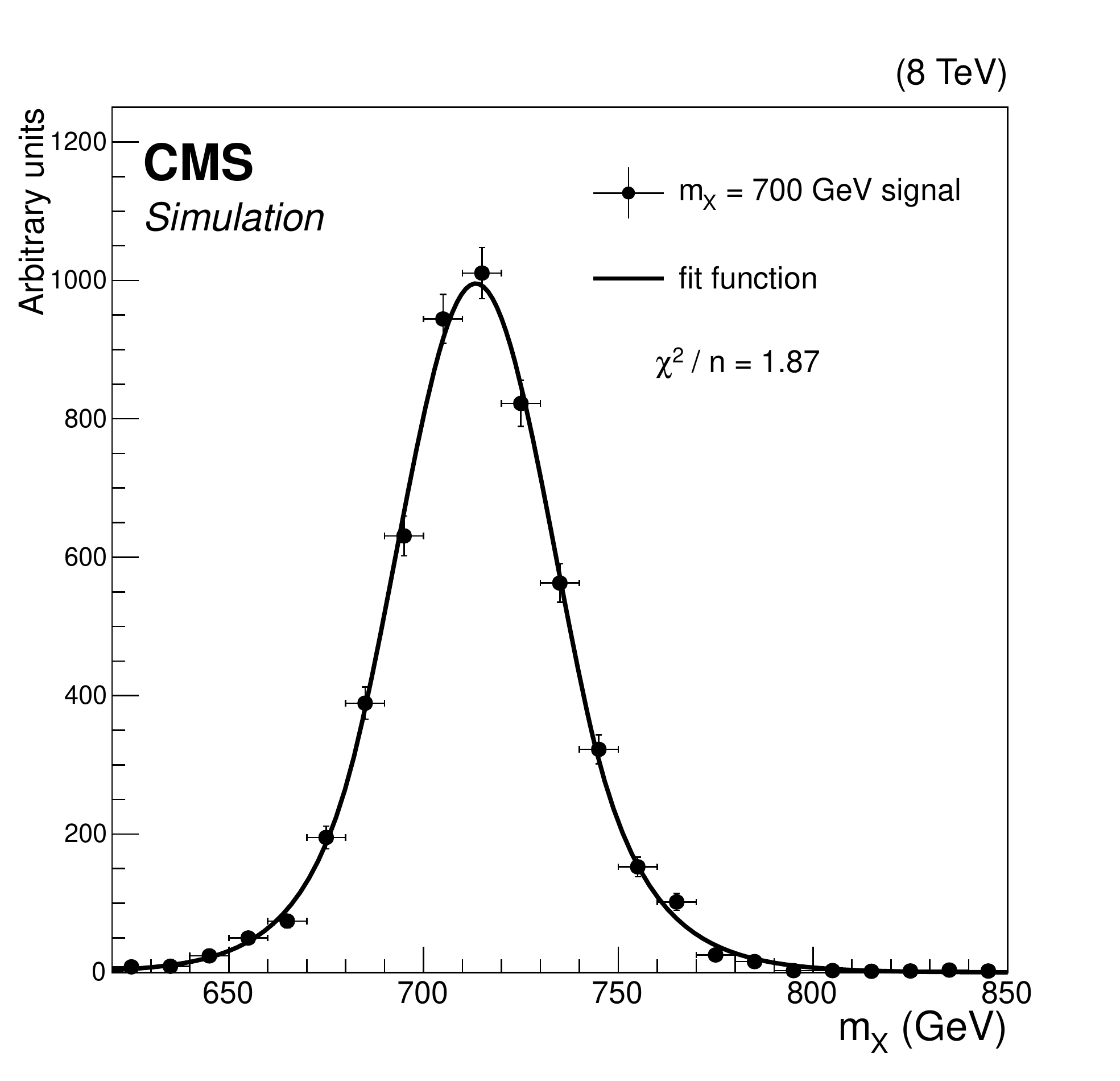}

\caption{A maximum likelihood fit to the $m_\cPX$ distribution of simulated signal events for the 700\GeV mass hypothesis. The distribution is fitted to a Gaussian core smoothly extended on both sides to exponential tails. Here $n$ is the number of degrees of freedom in the fit.}
\label{fig:FitSignal}
\end{figure}

For signal events, the aforementioned event selection criteria are expected to produce a sharp peak in the $m_\cPX$ distribution over a relatively featureless background from events arising from SM processes. The interference between SM background processes and the narrow resonant signal is expected to be negligible. To search for signal events at various mass hypotheses, we fit the $m_\cPX$ distribution in data events in the SR to a parametric model for the signal peak on top of parametric models appropriate for components of the SM background. This procedure is performed for the LMR, the MMR, and the HMR separately.

To improve the mass resolution of the signal \XHH resonance, we perform a fit that constrains the invariant masses of the Higgs boson candidates. In the fit, the momenta of the reconstructed b jets are allowed to float within their expected resolutions. Since the uncertainty in the reconstructed mass of the Higgs boson candidate due to the measurement of jet direction is smaller than that due to the measurement of jet energy, this constraint mainly affects the latter. This fit improves the invariant mass resolution of the reconstructed signal resonance by 20--40\%, depending on the mass hypothesis. Extensive tests in background-dominated control regions in data show that no artificial structures are introduced in the background mass distributions by this procedure.

We build the parametric model for each signal mass hypothesis by fitting the shape of the $m_\cPX$ distribution of simulated events that are accepted by the selection criteria and corrected for differences between data and simulation. A sum of two Gaussian functions, requiring five parameters, is used for the LMR fit to account for tails in the distribution from incorrect combinations of jets. In the MMR and the HMR, we fit a function with a Gaussian core smoothly extended on both sides to exponential tails, such that the function is continuous both in its value and its first derivative. This requires two parameters for the mean and width of the Gaussian function, and two other parameters for the exponential tails on both sides. An example of a parametric model for the MMR signal obtained through this procedure is shown in Fig.~\ref{fig:FitSignal}. While the model is constructed for the mass hypothesis of 700\GeV, its Gaussian core peaks at 714\GeV and has a width of 21\GeV. This mass shift is found to be linear in $m_\cPX$ and occurs due to the aforementioned constraint of jet momenta to $m_{\PH}$.

\begin{figure*}[hbtp]
\centering
\includegraphics[width=0.49\linewidth]{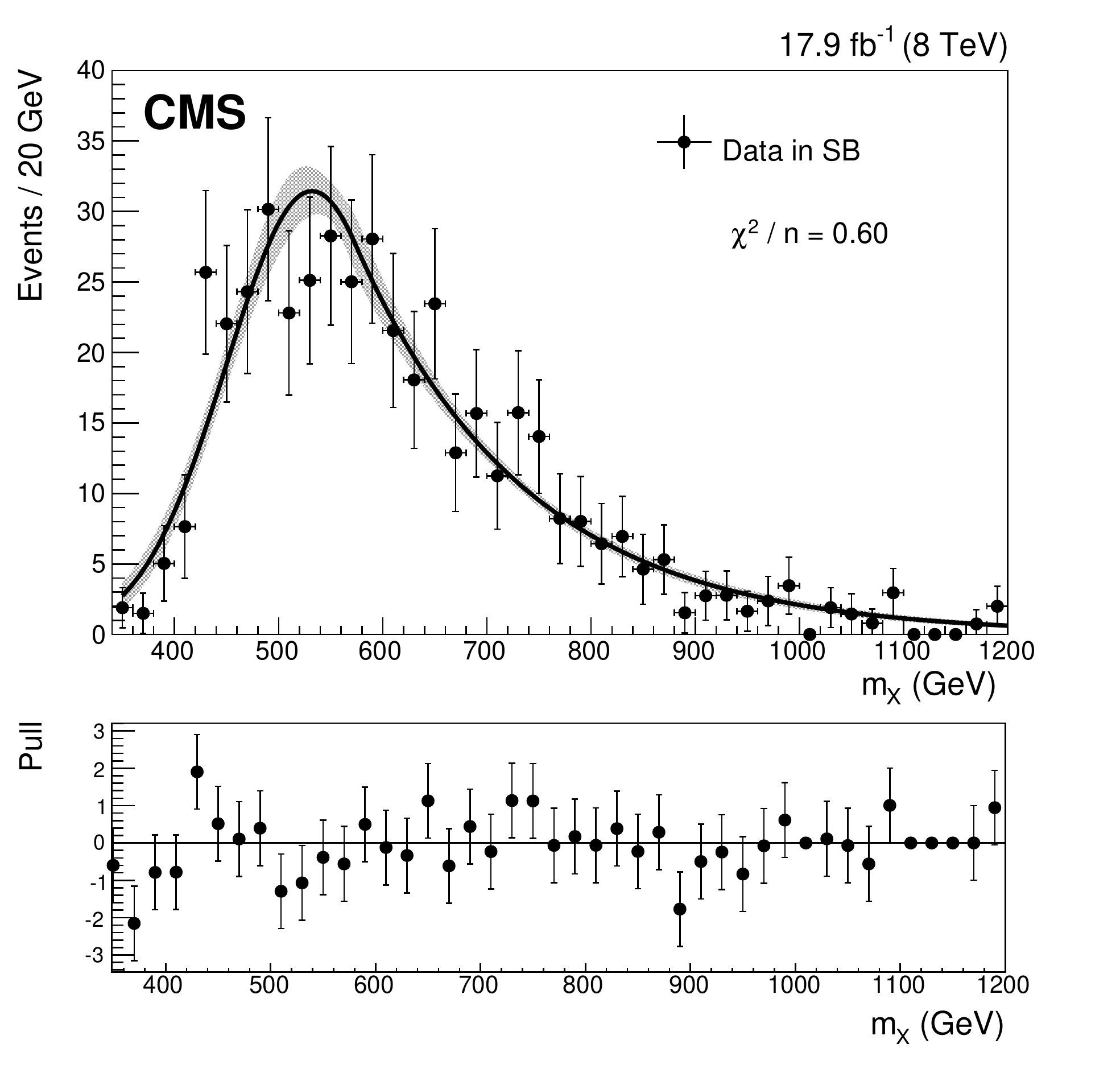}
\includegraphics[width=0.49\linewidth]{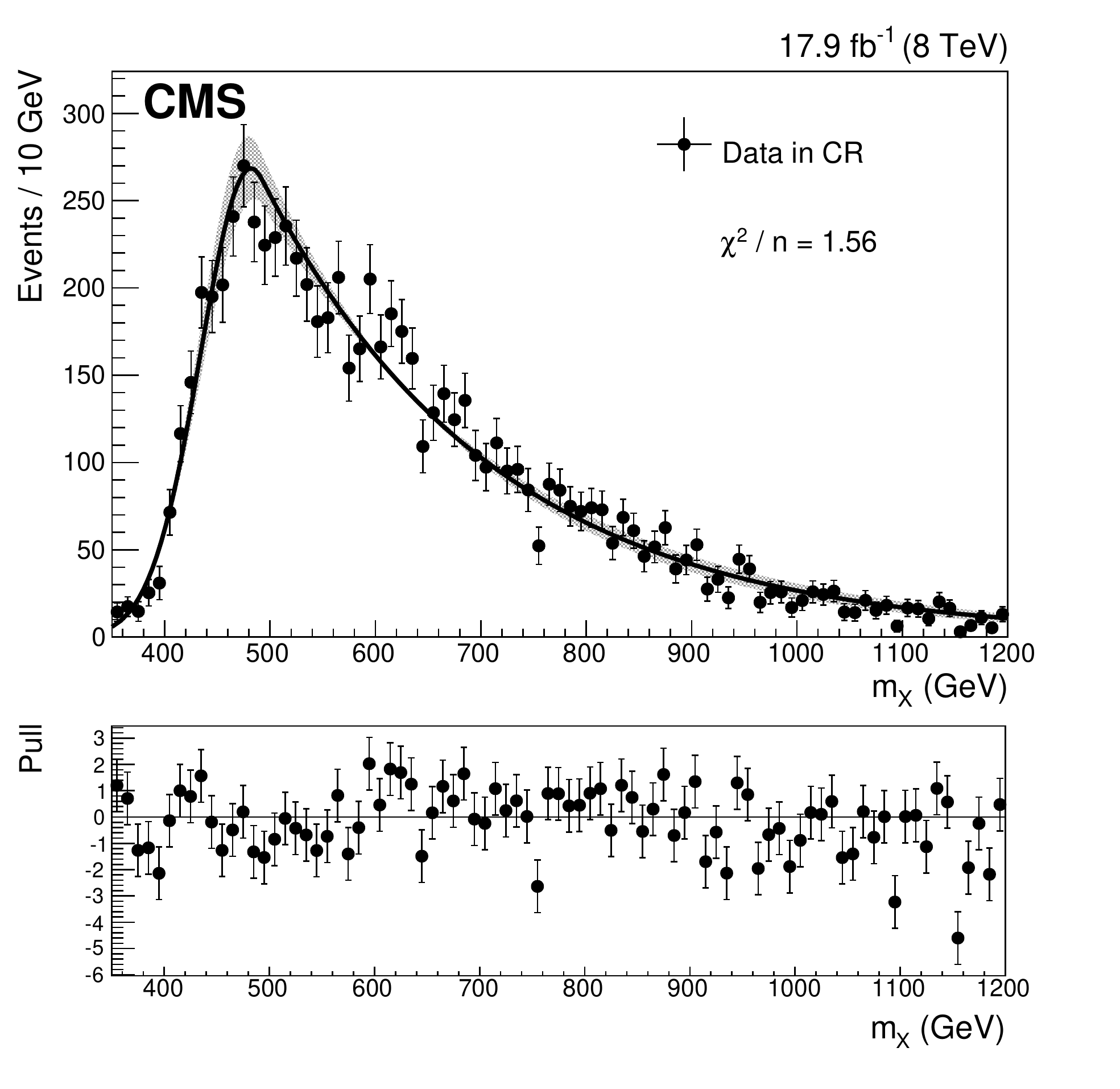}
\includegraphics[width=0.49\linewidth]{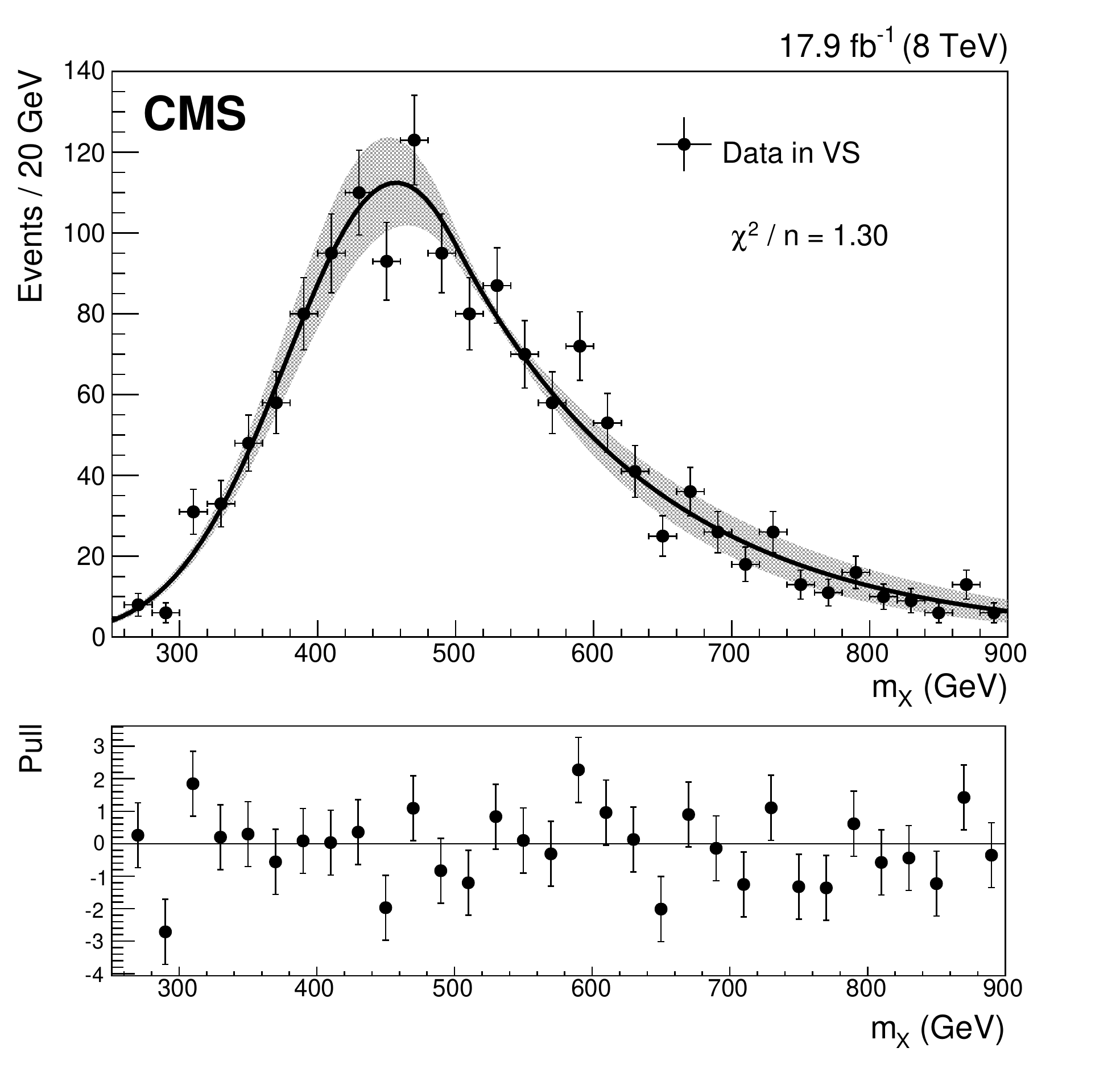}
\includegraphics[width=0.49\linewidth]{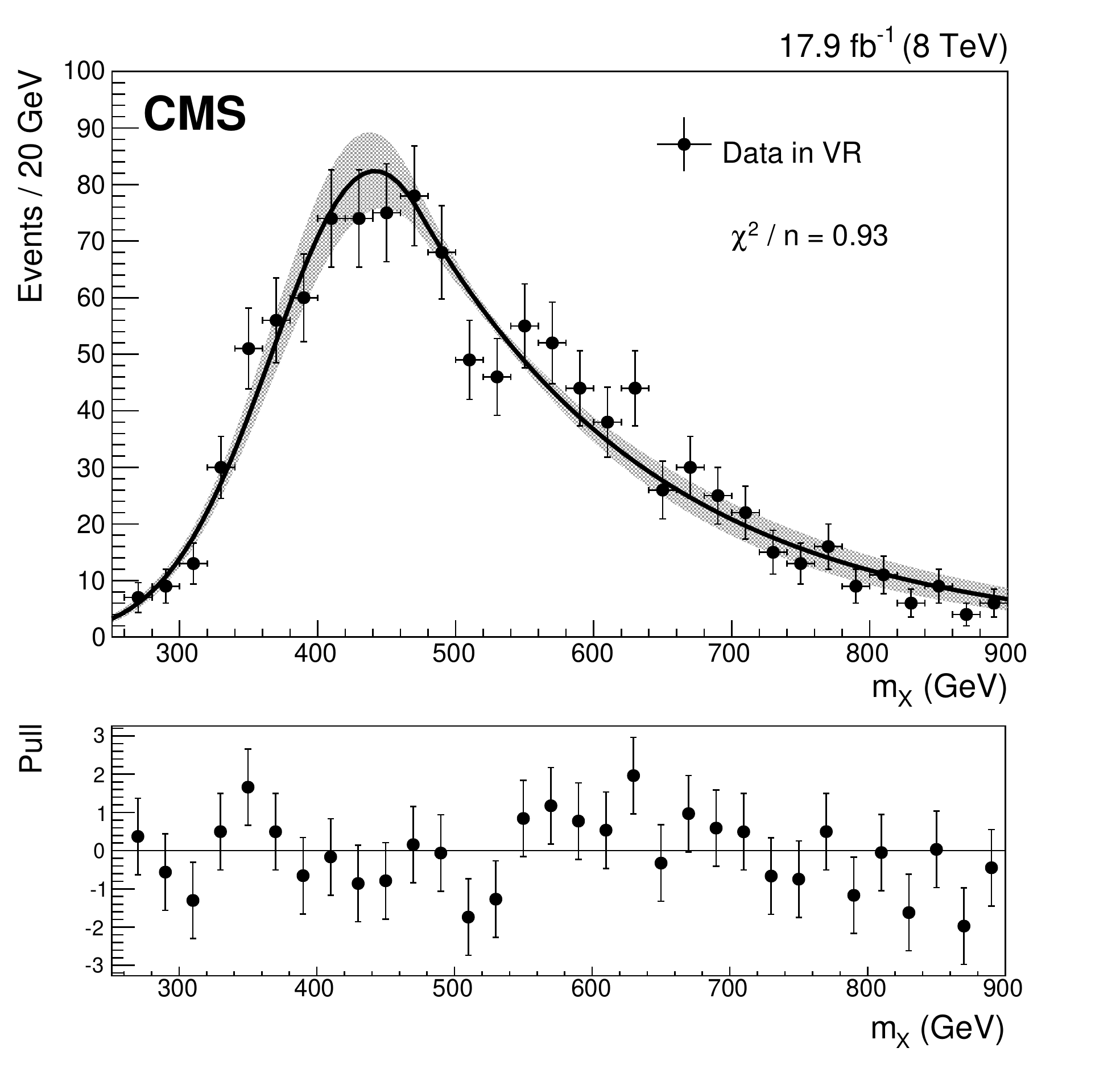}

\caption{The $m_\cPX$ distributions in data (after the \ttbar background has been subtracted) in the SB of the MMR (top left), the CR of the MMR (top right), the VS of the LMR (bottom left), and the VR of the LMR (bottom right). The distributions are fitted to the GaussExp parametric model. The shaded regions correspond to $\pm$1$\sigma$ variations of this fit. Here $n$ is the number of degrees of freedom in each fit. The pull, for a given bin, is defined as the number of data events minus the value of the fit model, divided by the uncertainty in the number of data events.}
\label{fig:MMMM}
\end{figure*}

\section{Background Modeling}

While the composition of background events in the SR is expected to be dominated by QCD multijet processes, we find through simulation that \ttbar production contributes approximately 22\%, 27\%, and 24\% in the LMR, MMR, and HMR, respectively. We also find that Z+jets, ZZ, and ZH processes contribute less than 1\% of the background and therefore neglect them in this analysis. The $m_\cPX$ distribution of these \ttbar events is found to be somewhat different in shape from that of QCD multijet events, and therefore we treat it as a distinct component of the background and model it with a parametric form. We obtain this parametric form by fitting the shape of the $m_\cPX$ distribution of simulated \ttbar events accepted by the event selection criteria to a function with a Gaussian core smoothly extended to an exponential tail on the high side. This function, henceforth referred to as GaussExp, is continuous in its value and its first derivative. It has two parameters for the mean and width of the Gaussian function and one parameter for the decay constant of the exponential tail. This model is normalized to a \ttbar cross section of 234\unit{pb}~\cite{Kidonakis:2012db}, and is allowed to float with a systematic uncertainty of 15\% in the final fit to account for theoretical and measurement uncertainties in our kinematically boosted region.

We use the GaussExp parametric model to fit the $m_\cPX$ distribution of the QCD multijet component of the background in the SR. With the SR kept blinded, we motivate and validate this choice of parametric model by the fact that it fits well the shape of the $m_\cPX$ QCD multijet background distributions in several different regions of the $(m_{{\PH}_1}, m_{{\PH}_2})$ plane depicted in Fig.~\ref{fig:Strategy} and described below. We do not aim to predict the parameters of the model in the SR from the other regions. These fits are performed for the LMR between 260 and 650\GeV, for the MMR between 400 and 900\GeV, and for the HMR between 600 and 1200\GeV. In each case the \ttbar contribution, as expected from simulation, is subtracted.

We define a sideband region (SB) to the SR as 17.5\GeV $<\sqrt{\smash[b]{\Delta m_{{\PH}_1}^{2} + \Delta m_{{\PH}_2}^{2}}}<35$\GeV and $\Delta m_{{\PH}_1}\,\Delta m_{{\PH}_2}<0$. For events in this region, the $m_\cPX$ distribution is expected to be kinematically similar to that for events in the SR, since in each of the sidebands one of the reconstructed Higgs boson masses is slightly higher in value than for events in the SR while the other is slightly lower. As an example, Fig.~\ref{fig:MMMM} top left shows the fit performed for events in the SB passing the HMR selection. Another set of events that pass the kinematic requirements of the event selection criteria in the SR region of the $(m_{{\PH}_1}, m_{{\PH}_2})$ plane but required to have one of the four jets not be b-tagged is selected to further test the applicability of the GaussExp model in describing the $m_\cPX$ distribution of the QCD multijet background in a different but kinematically similar region. This is called the data Control Region (CR), and the fit for these events, that would have otherwise passed the HMR selection, is also shown in Fig.~\ref{fig:MMMM} on the top-right. In both cases, the goodness of the fit, characterized by the $\chi^2$ per degree of freedom, is found to be reasonable.

These two cases already lend significant confidence to the choice of the GaussExp parametric model for the SR. However, we carry out further checks in neighboring validation regions (VR) with a corresponding sideband (VS) that are defined similarly to the SR and SB regions but with $m_{{\PH}_1}=m_{{\PH}_2}$ centered at different values. The good fits for the $m_\cPX$ distributions in these regions not only demonstrate the applicability of the GaussExp model to describe these kinematically distinct QCD multijet events, but also that events in the VR are in fact kinematically similar to those in the VS. As examples, Fig.~\ref{fig:MMMM} bottom-left and bottom-right plots show the results of these fits for the LMR selection for the VS and the VR, respectively, both centered at $m_{{\PH}_1}=m_{{\PH}_2} =90$\GeV. We obtain similar results for the VR centered at $m_{{\PH}_1} = m_{{\PH}_2} = 107.5$, 142.5 and 160\GeV.

While the GaussExp function fits well the $m_\cPX$ distribution from QCD multijet events in all these distinct regions and therefore can be expected to be a good approximation of the parametric form of the true parent distribution for events in the SR, other similar parametric models could be chosen instead. Therefore, a systematic uncertainty associated with the choice of this parametric model is evaluated by assuming a 7${th}$ order Bernstein polynomial, which also fits the $m_\cPX$ distribution well in the SB, to be the true distribution. Pseudo-datasets are generated from this polynomial function and fitted with the GaussExp function as well as other polynomial functions to compute biases in the reconstructed signal strength. This procedure is performed for each mass hypothesis. These biases are found to be of the order of 100\unit{fb} for the LMR, 10\unit{fb} for the MMR, and 20\unit{fb} for the HMR. We account for this bias as a signal-shaped systematic uncertainty in the background model with normalization centered at zero and a Gaussian uncertainty with standard deviation equal to the bias.

\begin{table*}[htb]
\centering
\topcaption{Relative systematic uncertainties in the selection efficiencies for signal and \ttbar events in the LMR, the MMR, and the HMR.}
\label{tab:Systematics}
\begin{tabular}{cc.@{\hspace{1em}}cc.@{\hspace{1em}}cc.}
\hline
\multirow{2}{20ex}{\centering\arraybackslash Source of systematic uncertainty}               & \multicolumn{2}{c}{Impact in LMR (\%)} && \multicolumn{2}{c}{Impact in MMR (\%)} && \multicolumn{2}{c}{Impact in HMR (\%)} \\\cline{2-3}\cline{5-6}\cline{8-9}
 & Signal      & \multicolumn{1}{c}{\ttbar}    && Signal     & \multicolumn{1}{c}{\ttbar}   && Signal     & \multicolumn{1}{c}{\ttbar} \rule{0pt}{2.5ex}  \\
\hline
Jet energy scale         & 0.1--0.2   & 0.8       && 0.0--0.2  & 0.1      && 0.1--0.3 & 1.9   \\
Jet energy resolution    & 2.4--7.0   & 2.7       && 5.5--7.0  & 2.2      && 4.9--5.3 & 7.2   \\
b-tagging scale factor   & 12.7       & 12.7      && 12.7      & 12.7     && 12.7     & 12.7  \\
Trigger scale factor     & 6.0--18.8  & 9.1       && 6.1--8.0  & 7.2      && 7.4--7.9 & 7.2   \\
\hline
\end{tabular}
\end{table*}

\section{Systematic Uncertainties}

Sources of systematic uncertainties that affect the selection efficiencies for signal and \ttbar events are listed in Table~\ref{tab:Systematics} and described below. We vary the jet energy scale~\cite{cmsJEC} by its uncertainty as a function of jet \pt and $\eta$ and find that this affects signal efficiencies by a relative factor of up to 0.2\% and \ttbar efficiencies by up to 0.8\%. We evaluate the effect of uncertainty in the jet energy resolution by varying the jet energies according to the measured uncertainty. This is found to affect signal efficiencies by 2--7\%, and \ttbar efficiencies by 2--3\%. These uncertainties affect not only the normalizations but also the parameters of the signal and \ttbar models, and are taken into account as nuisance parameters in the final fit.

The trigger efficiencies for signal and \ttbar events are evaluated approximately by passing generated events through a trigger simulation. We then correct these efficiencies for differences between simulation and data by computing the difference in a \ttbar-enriched control region obtained using a trigger that requires at least one muon with $\pt>24$\GeV. Further, event selection criteria requiring at least one muon with $\pt>40$\GeV and at least four jets in the central region of the detector with $\pt>40$\GeV are applied. The data-to-simulation correction factor is characterized by the \pt and CMVA discriminants of the relevant jets. Uncertainties in this factor impact signal efficiencies by 6--18\%, and \ttbar efficiencies by 7--9\%.

The b-tagging efficiencies of the CMVA algorithm for signal and \ttbar events are also evaluated approximately through simulation and then corrected by a data-to-simulation comparison. The comparison is performed in the same \ttbar-enriched control region as the calculation of the trigger efficiency. The correction factor for the b-tagging efficiencies is consistent with unity. The uncertainty in this factor for four b jets is evaluated to be 12.7\%.

Additionally, the yields of signal events for a given production cross section and \ttbar events are both affected by a 2.6\% uncertainty in the measurement of the integrated luminosity~\cite{CMS-PAS-LUM-13-001}.

\section{Results}

\begin{figure*}[htbp]
\centering
\includegraphics[width=\cmsFigWidthThree]{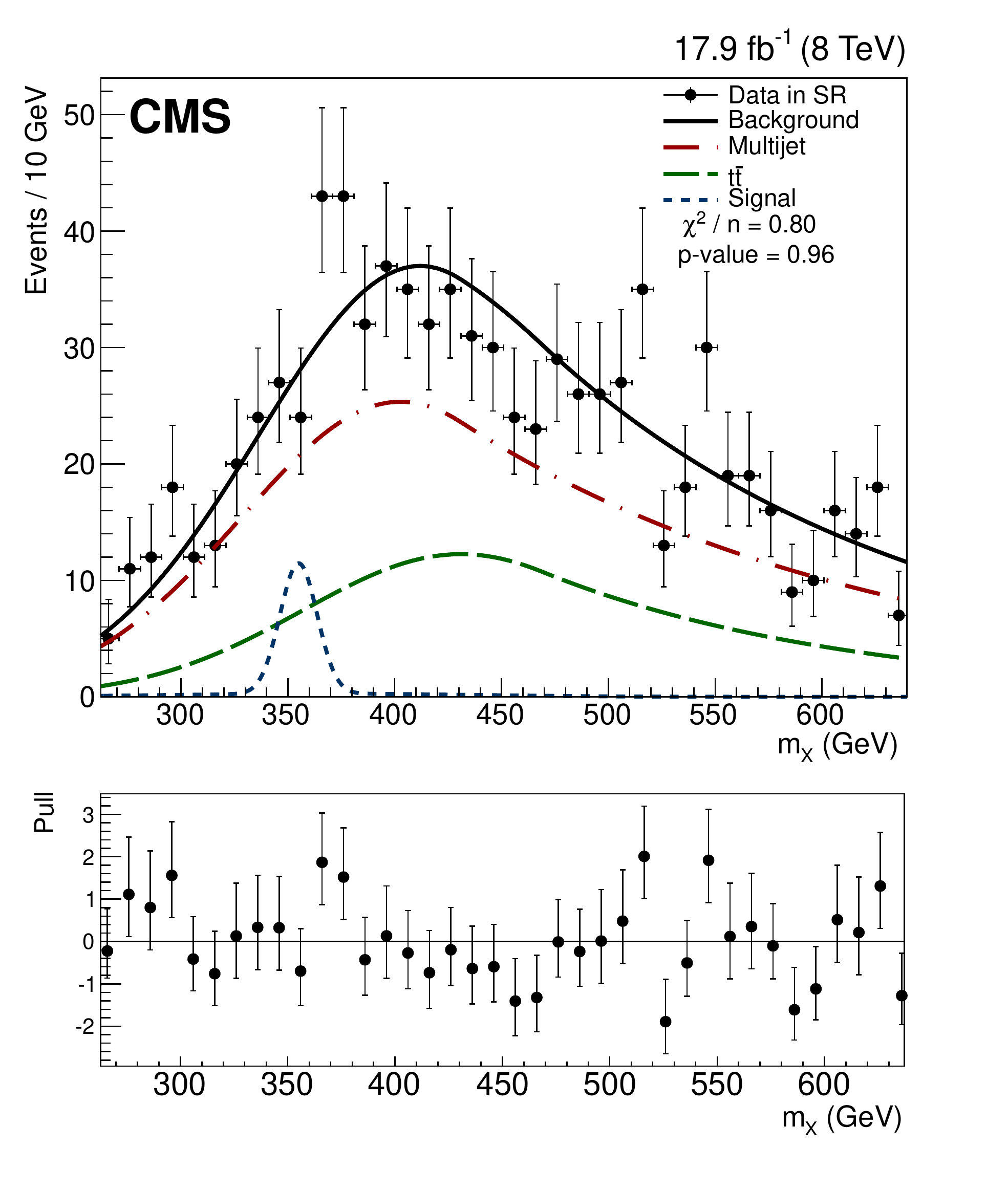}
\includegraphics[width=\cmsFigWidthThree]{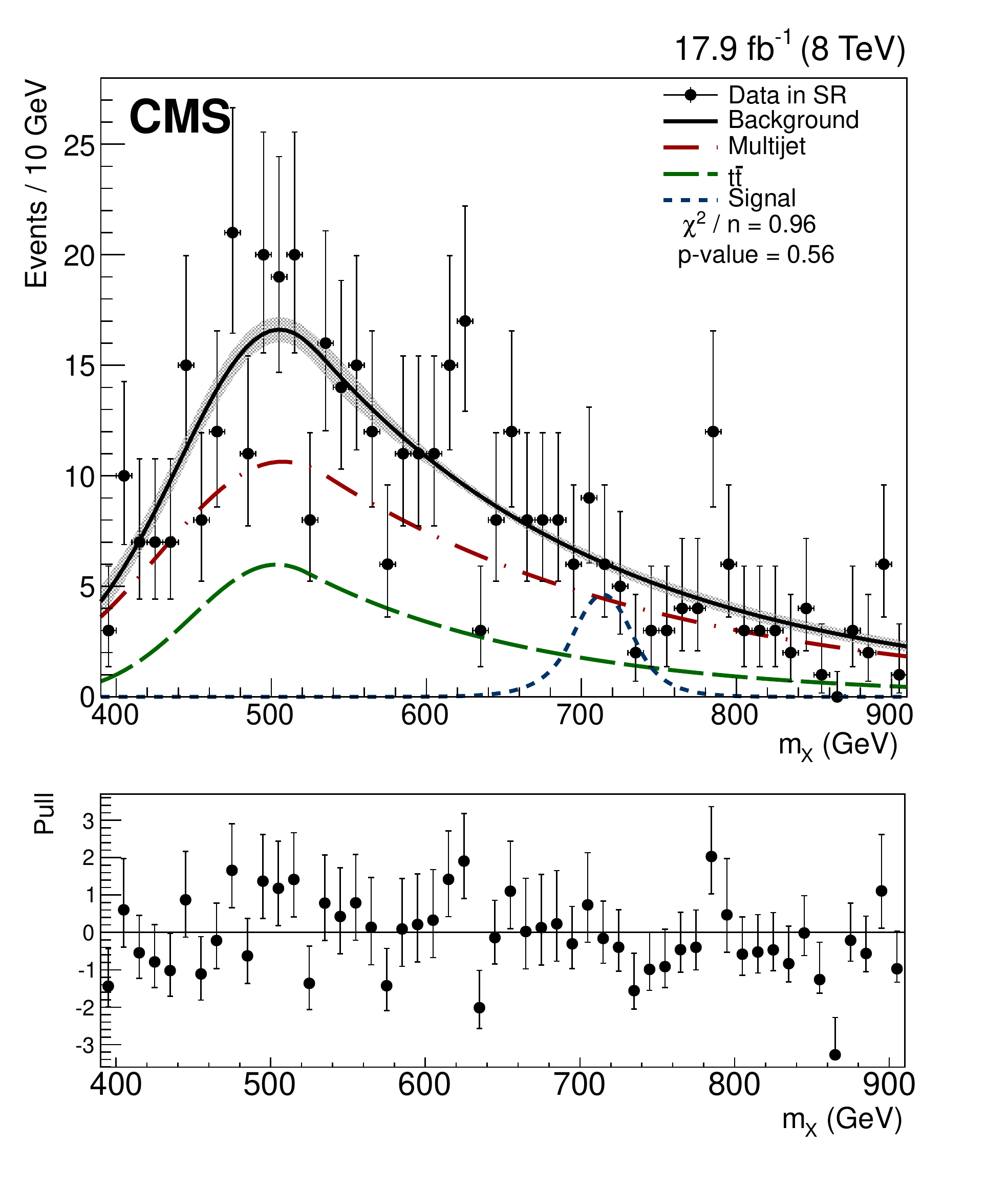}
\includegraphics[width=\cmsFigWidthThree]{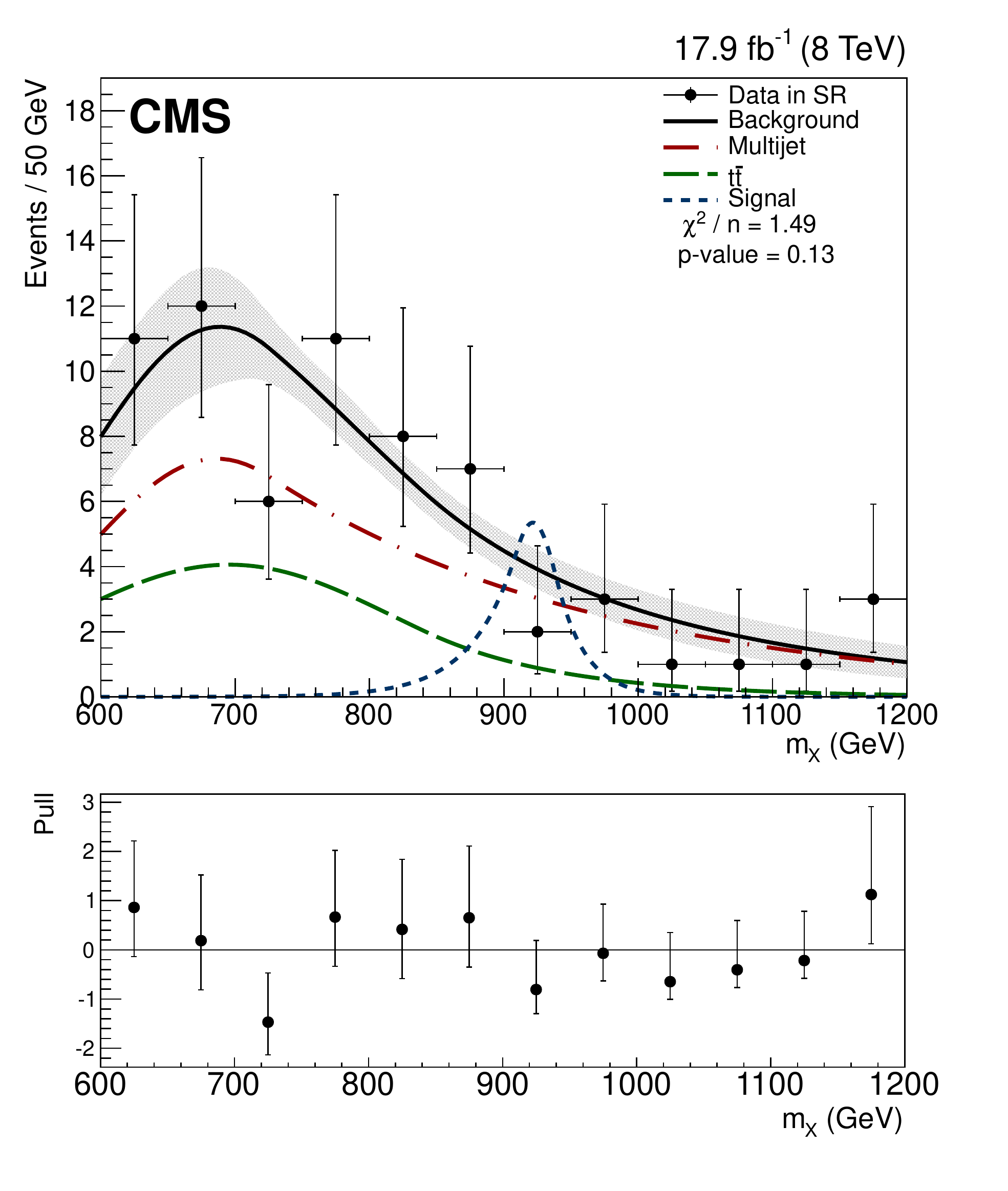}
\caption{The $m_{\mathrm{X}}$ distribution in data in the SR between 260 and 650\GeV of the LMR (top left), between 400 and 900\GeV of the MMR (top right), and between 600 and 1200\GeV in the HMR (bottom). All distributions are fitted to the background-only hypothesis for illustration, showing the relative contributions of the QCD multijet (dashed-dotted red) and \ttbar (dashed green) processes. The pull, for a given bin, is defined as the number of data events minus the value of the background-only fit, divided by the uncertainty in the number of data events. Also for illustration, we overlay the signal models of the spin-0 resonance (dotted blue) corresponding to mass hypotheses and production cross sections of 350\GeV and 653\unit{fb} for the LMR, 700\GeV and 17.6\unit{fb} for the MMR, and 900\GeV and 8.1\unit{fb} for the HMR. These cross sections correspond to the observed upper limits, which are computed for signal mass hypotheses from 270 to 450\GeV in the LMR, from 450 to 730\GeV in the MMR, and from 730 to 1100\GeV in the HMR.
}
\label{fig:unblinded}
\end{figure*}

\begin{figure}[htbp]
\centering
\includegraphics[width=0.49\textwidth]{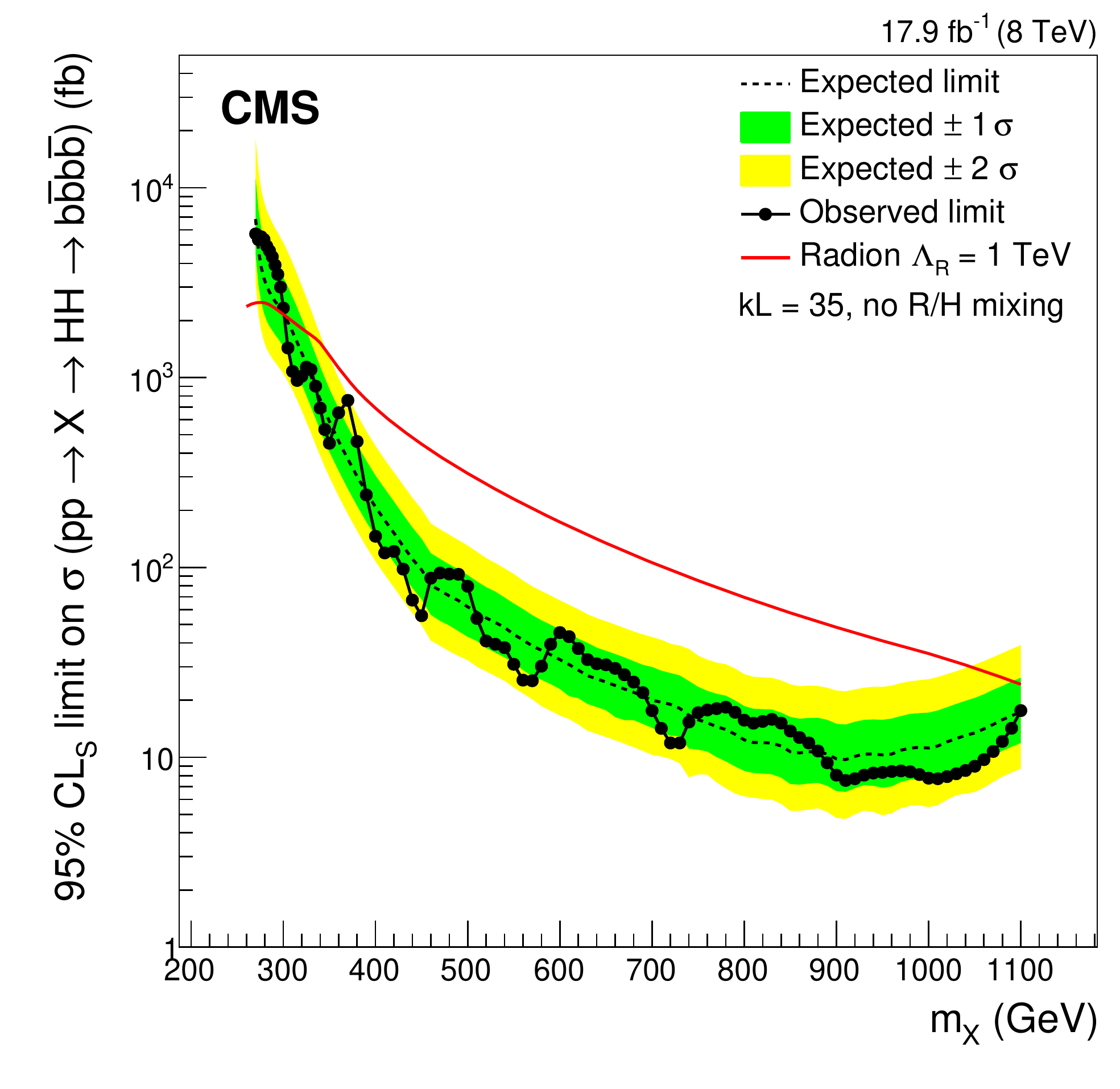}
\includegraphics[width=0.49\textwidth]{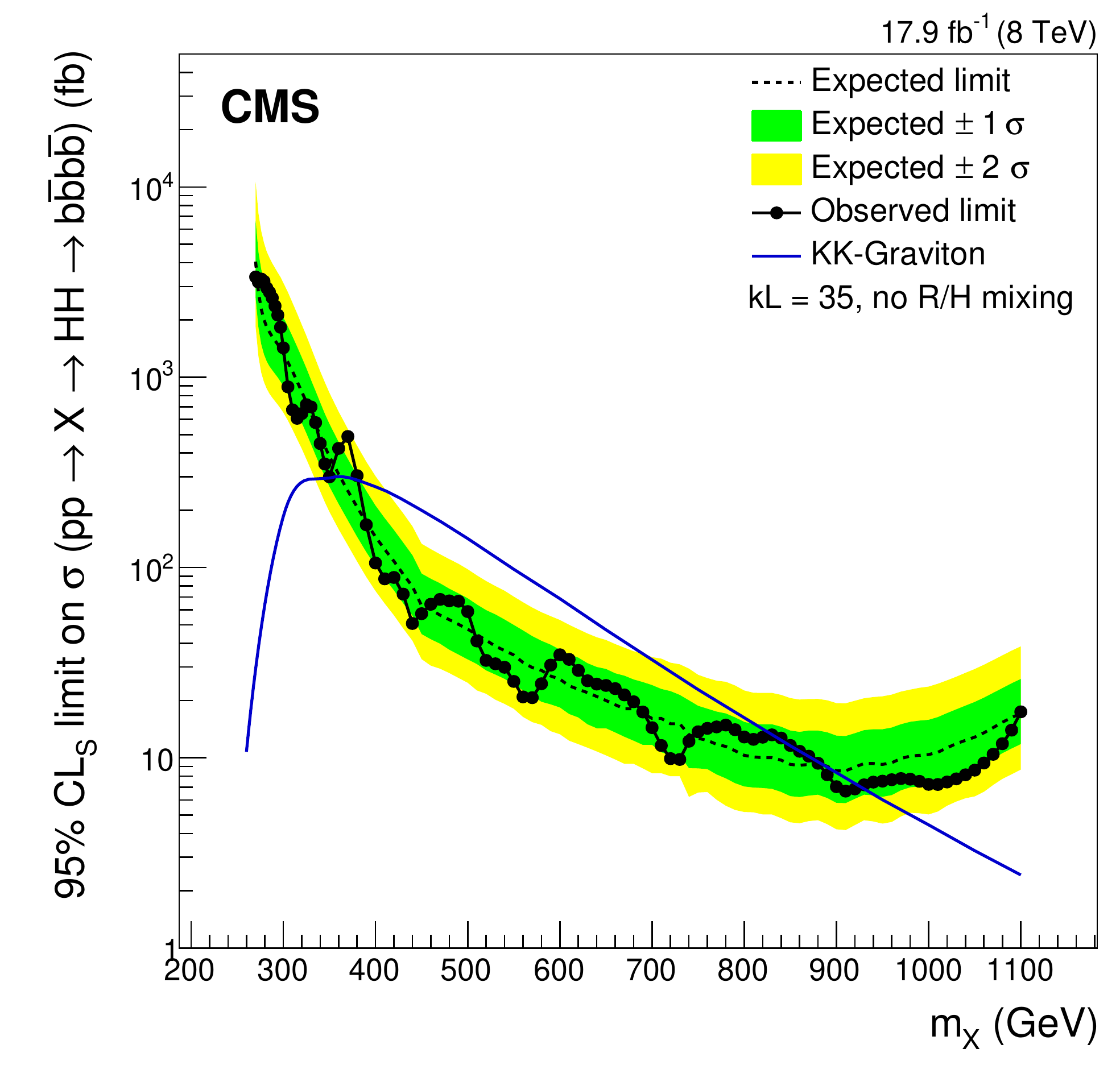}

\caption{The observed and expected upper limits on the cross section for \ppXHbbHbb at a 95\% confidence level, where the resonance X has spin-0 (\cmsLeft) and spin-2 (\cmsRight). The theoretical cross section for the RS1 radion, with $\Lambda_{R} = $1\TeV, $kL = 35$, and no radion-Higgs boson mixing, decaying to four b jets via Higgs bosons is overlaid on the left plot. The theoretical cross section for the first excitation of the KK-graviton for the same parameters is overlaid on the \cmsRight plot.}
\label{fig:UpperLimit_Combined}
\end{figure}

The $m_\cPX$ distribution that we observe in data within the SR, along with a binned maximum-likelihood fit with the aforementioned parametric background models, are shown in Fig.~\ref{fig:unblinded}. We compute the observed and expected upper limits on the cross section for \ppXHbbHbb at a 95\% confidence level (CL) using the modified frequentist CL$_\mathrm{S}$ method~\cite{Read, CMS-NOTE-2011-005} by fitting the data with the parametric signal, \ttbar, and QCD multijet models. This is done separately in the disjoint ranges of $m_\cPX$ for the individual regions described in Section~\ref{sec:EventSelection}, and the limits are presented together in Fig.~\ref{fig:UpperLimit_Combined}. These limits are shown for the spin-0 resonance on the left, and the spin-2 resonance on the right. The green (dark) and yellow (light) bands respectively represent the 1$\sigma$ and 2$\sigma$ confidence intervals around the expected limits. The observed upper limits lie within 2$\sigma$ of the expected upper limits, and thus we conclude that there is no significant deviation from the background-only hypothesis.

The theoretical cross section for the production via gluon fusion of a radion that decays to a pair of Higgs bosons~\cite{Barger:2011qn} that each in turn decays to a \bbbar pair with a branching fraction of 58\%~\cite{Dittmaier:2011ti} is calculated using \MADGRAPH 5.1~\cite{RadionMC} and superimposed on the experimental cross section limit for the spin-0 resonance in the plot on the left. In this calculation, the correction factor used to account for next-to-leading-order effects for electroweak couplings~\cite{Actis:2008ug} and next-to-next-to-leading-order effects for QCD couplings~\cite{Heinemeyer} is identical to that used for Higgs boson production through gluon fusion. The warped extra dimension scenario for this radion has the product of the curvature, $k$, and half the circumference of the extra dimension, $L$, set to 35, a radion decay constant of $\Lambda_{R} = 1$\TeV, and no radion-Higgs boson mixing. The theoretical cross section for the radion has an uncertainty of approximately 15\% that is not used to compute the experimental limits on spin-0 resonance production shown in Fig.~\ref{fig:UpperLimit_Combined}. Masses for the radion between 300 and 1100\GeV are excluded at a 95\% CL. A similarly calculated theoretical cross section for the KK graviton as the resonance X, in the same warped extra dimension scenario, is overlaid on the limit for the spin-2 resonance in the plot on the right. Masses for such a graviton are excluded at a 95\% CL between 380 and 830\GeV.

\section{Summary}

We have presented a model-independent search by the CMS experiment at the LHC for a narrow resonance produced in proton-proton collisions at $\sqrt{s} =8$\TeV and decaying to a pair of 125\GeV Higgs bosons that in turn each decays into a bottom quark-antiquark pair. The analyzed data correspond to an integrated luminosity of 17.9 \fbinv. No evidence for a signal is observed. Upper limits at a 95\% CL on the production cross section for such spin-0 and spin-2 resonances, in the mass range from 270 to 1100\GeV, are reported. Using these results, a radion with decay constant of 1\TeV and mass from 300 to 1100\GeV, and a Kaluza--Klein graviton with mass from 380 to 830\GeV are excluded at a 95\% confidence level.

\begin{acknowledgments}
We congratulate our colleagues in the CERN accelerator departments for the excellent performance of the LHC and thank the technical and administrative staffs at CERN and at other CMS institutes for their contributions to the success of the CMS effort. In addition, we gratefully acknowledge the computing centers and personnel of the Worldwide LHC Computing Grid for delivering so effectively the computing infrastructure essential to our analyses. Finally, we acknowledge the enduring support for the construction and operation of the LHC and the CMS detector provided by the following funding agencies: BMWFW and FWF (Austria); FNRS and FWO (Belgium); CNPq, CAPES, FAPERJ, and FAPESP (Brazil); MES (Bulgaria); CERN; CAS, MoST, and NSFC (China); COLCIENCIAS (Colombia); MSES and CSF (Croatia); RPF (Cyprus); MoER, ERC IUT and ERDF (Estonia); Academy of Finland, MEC, and HIP (Finland); CEA and CNRS/IN2P3 (France); BMBF, DFG, and HGF (Germany); GSRT (Greece); OTKA and NIH (Hungary); DAE and DST (India); IPM (Iran); SFI (Ireland); INFN (Italy); MSIP and NRF (Republic of Korea); LAS (Lithuania); MOE and UM (Malaysia); CINVESTAV, CONACYT, SEP, and UASLP-FAI (Mexico); MBIE (New Zealand); PAEC (Pakistan); MSHE and NSC (Poland); FCT (Portugal); JINR (Dubna); MON, RosAtom, RAS and RFBR (Russia); MESTD (Serbia); SEIDI and CPAN (Spain); Swiss Funding Agencies (Switzerland); MST (Taipei); ThEPCenter, IPST, STAR and NSTDA (Thailand); TUBITAK and TAEK (Turkey); NASU and SFFR (Ukraine); STFC (United Kingdom); DOE and NSF (USA).

Individuals have received support from the Marie-Curie program and the European Research Council and EPLANET (European Union); the Leventis Foundation; the A. P. Sloan Foundation; the Alexander von Humboldt Foundation; the Belgian Federal Science Policy Office; the Fonds pour la Formation \`a la Recherche dans l'Industrie et dans l'Agriculture (FRIA-Belgium); the Agentschap voor Innovatie door Wetenschap en Technologie (IWT-Belgium); the Ministry of Education, Youth and Sports (MEYS) of the Czech Republic; the Council of Science and Industrial Research, India; the HOMING PLUS program of the Foundation for Polish Science, cofinanced from European Union, Regional Development Fund; the Compagnia di San Paolo (Torino); the Consorzio per la Fisica (Trieste); MIUR project 20108T4XTM (Italy); the Thalis and Aristeia programs cofinanced by EU-ESF and the Greek NSRF; and the National Priorities Research Program by Qatar National Research Fund.
\end{acknowledgments}
\bibliography{auto_generated}
\cleardoublepage \appendix\section{The CMS Collaboration \label{app:collab}}\begin{sloppypar}\hyphenpenalty=5000\widowpenalty=500\clubpenalty=5000\input{HIG-14-013-authorlist.tex}\end{sloppypar}
\end{document}

%% file: HIG-14-013-authorlist.tex
\textbf{Yerevan Physics Institute,  Yerevan,  Armenia}\\*[0pt]
V.~Khachatryan, A.M.~Sirunyan, A.~Tumasyan
\vskip\cmsinstskip
\textbf{Institut f\"{u}r Hochenergiephysik der OeAW,  Wien,  Austria}\\*[0pt]
W.~Adam, T.~Bergauer, M.~Dragicevic, J.~Er\"{o}, M.~Friedl, R.~Fr\"{u}hwirth\cmsAuthorMark{1}, V.M.~Ghete, C.~Hartl, N.~H\"{o}rmann, J.~Hrubec, M.~Jeitler\cmsAuthorMark{1}, W.~Kiesenhofer, V.~Kn\"{u}nz, M.~Krammer\cmsAuthorMark{1}, I.~Kr\"{a}tschmer, D.~Liko, I.~Mikulec, D.~Rabady\cmsAuthorMark{2}, B.~Rahbaran, H.~Rohringer, R.~Sch\"{o}fbeck, J.~Strauss, W.~Treberer-Treberspurg, W.~Waltenberger, C.-E.~Wulz\cmsAuthorMark{1}
% ...
\vskip\cmsinstskip
% ...
1:~~Also at Vienna University of Technology, Vienna, Austria\\
2:~~Also at CERN, European Organization for Nuclear Research, Geneva, Switzerland\\
% ...
